\documentclass{emulateapj}
\usepackage{ae} 
\usepackage[T1]{fontenc}
\usepackage[ansinew]{inputenc}
\usepackage{amsmath}
\usepackage{amssymb}
\usepackage{amsfonts}
\usepackage{natbib}

\newcommand{\ka}[1]{\langle #1\rangle}
\newcommand{\kB}{k_{\text{B}}}

\slugcomment{Submitted to The Astrophysical Journal}

\shorttitle{Energetics and partition function of H$_3^+$}

\begin{document}

\title{Energetics and partition function of H$_3^+$ molecular ion}
\author{Ilkka Kylänpää\altaffilmark{1} and Tapio T. Rantala\altaffilmark{2}}
\affil{Department of Physics, Tampere 
University of Technology, P.O. Box 692, FI-33101 Tampere, Finland}

\altaffiltext{1}{email: Ilkka.Kylanpaa@tut.fi} 
\altaffiltext{2}{email: Tapio.Rantala@tut.fi}

\begin{abstract}
Full $NVT$ quantum statistics of the H$_3^+$ ion is simulated at low
densities using the path integral Monte Carlo approach.  For the first
time, the molecular total energy, partition function, free energy,
entropy and heat capacity are evaluated in temperatures relevant for
planetary atmospheric physics.
Temperature and density dependent dissociation recombination reaction
balance of the molecule and its fragments above $4000$ K is described,
and also, the density dependence of thermal ionization above $10 000$
K is demonstrated.
We introduce a new well-behaving analytical model for the molecular
partition function of the H$_3^+$ ion for the temperature range below
dissociation and fit the parameters to the energetics from our
simulations.
The approach presented here can be regarded as an extension of the
traditional {\it ab initio} quantum chemistry beyond the
Born--Oppenheimer approximation to description of nonadiabatic
phenomena, and even further, account of nuclear quantum dynamics.
\end{abstract}



\maketitle 

\section{Introduction}

The H$_3^+$ molecular ion has been a subject of a number of
theoretical and experimental studies since its first experimental detection
\citep{Thomson11}. Because of its rapid formation through the
exothermic reaction ($\Delta E\approx -1.7$ eV)
\begin{align}
\rm{H}_2 + \rm{H}_2^+ \rightarrow \rm{H}_3^+ + \rm{H} 
\end{align}
the H$_3^+$ ion is expected in any active environment containing
molecular hydrogen \citep{Neale95ApJ}, and thus, it is encountered,
e.g.~in the studies of the giant planets
\citep{Miller08ApJ,Koskinen09ApJ}. 

In planetary atmospheric physics, importance of the
H$_3^+$ ion lies in its capability to act as a cooling agent
via infrared radiation
\citep{Neale96ApJ,Tennyson04ApJ,Koskinen07ApJ}.
The atmospheric models taking into account this cooling
are commonly based on the high temperature molecular partition function
of the H$_3^+$ ion \citep{Neale95ApJ}.
Evaluation of the partition function faces, however, a few challenges
of which the first one is finding a good approximation to
the infinite summation over all rovibrational quantum states with accurate
enough energies  \citep{Neale95ApJ}.
This has usually been worked out with calculations of a
finite number of states from, e.g.~a semi-empirical potential
energy surface \citep{Tennyson95jcp}.

The next challenge comes with the changing geometry of the H$_3^+$ ion
at finite temperature.  The rovibrational model needs to be extended
for calculations of correct energetics for the emerging linear
geometry of the weakly bound molecule \citep{Neale96ApJ}.

Finally, at about the same temperature, where linear geometries
start contributing, the molecule may also dissociate to its fragments,
and in fact, the equilibrium
\begin{align}
\label{fragments}
\text{H}_3^+ & \leftrightarrow
\text{H}_2 + \text{p}^+ \nonumber \\
&\leftrightarrow
\text{H}_2^+ + \text{H}  \nonumber \\
&\leftrightarrow
2\text{H} + \text{p}^+  \nonumber \\
&\leftrightarrow
\text{H} + 2\text{p}^+ + \text{e}^-  \nonumber \\
&\leftrightarrow
3\text{p}^+ + 2\text{e}^- 
\end{align}
needs to be considered above about $4000$ K,
the balance depending strongly on both the temperature
and the density.

This brings forth two questions, at the least. First, how relevant it
is to consider the molecular energetics and related partition function
at temperatures where the molecule has dissociated and appears in form
of fragments of the equilibrium reaction, Eq.~\eqref{fragments}, only.
Secondly, the balance of the equilibrium reaction may be strongly
affected, not only by the density, but also by the environment
including the neutralizing negative counterparts of the positive
H$_3^+$.  Thus, the thermal dissociation--recombination balance above
about $4000$ K gives rise to problems, which have not been taken into
account in this context, yet.

In this study, using the path integral quantum Monte Carlo (PIMC)
method we have carried out the first simulations of the full quantum
statistics of the H$_3^+$ ion, described by Eq.~\eqref{fragments}, at
low densities and high temperatures ranging from $160$ K up to about
$15000$ K.  PIMC is the method to meet the above challenges: we need
not make any approximations or restrictions in the summing over
states, geometries or quantum description of dynamics.  The finite
temperature is inherent in the PIMC approach and the Coulomb many-body
treatment of the particle interactions is exact.  The PIMC method is
computationally expensive, but feasible for small enough systems.
\citep{Ceperley95,Pierce99,Kwon99,Knoll00,Cuervo06,Kylanpaa09pra}.

The conventional quantum chemical {\it ab initio} description of the
H$_3^+$ ion emerges as the zero Kelvin extrapolate from the PIMC
simulations as we have shown earlier \citep{Kylanpaa10jcp}.  There, we
evaluated the differences between three models for the description of
the nuclear dynamics: Born--Oppenheimer approximation, nuclei in
thermal motion and nuclei with both thermal and quantum dynamics.  At
low temperatures the necessity of the fully quantum mechanical
approach for all five particles was established.

In the next section we present the essential details of the Feynman
path integral quantum statistical approach, numerical simulation
method and the model of the H$_3^+$ ion.  In section 3 we present and
analyze the energetics, partition function and other thermodynamic
functions of the system using analytical forms where pertinent.  In
the last section the conclusions are given.

\section{Method and Models}

According to the Feynman path integral formulation of the quantum statistical
mechanics \citep{Fey72} the partition function of interacting
distinguishable particles is given by the trace of the density matrix
$\hat{\rho}(\beta) = e^{-\beta\hat{H}}$ as
\begin{align}
Z
\!\!= \!\!\text{Tr}\hat{\rho}(\beta)
\!\!= \!\!\!\int \!\!\text{d} R_{0}\text{d} R_{1}\!\ldots
\text{d} R_{M-1}\!\!\prod_{i = 0}^{M-1}\!\!e^{-S(R_{i},R_{i+1};\tau)},\nonumber
\end{align}
where the action $S(R_{i},R_{i+1};\tau)$ is taken over the path $R_{i}
\rightarrow R_{i+1}$ in imaginary time $\tau = \beta/M$, where $\beta
= 1/k_{\text{B}}T$ and $M$ is called the Trotter number.  The trace
implies a closed path ($R_{M}=R_{0}$).

For simulation we use the pair approximation in the action
\citep{Storer68,Ceperley95} for the Coulomb interaction of charges.
This is exact in the limit $M\rightarrow\infty$, however, chemical
accuracy is reached with sufficiently large $M$, i.e.~small enough
$\tau$.  Sampling in the configuration space $\{ R_i \}_i^\infty$ in $NVT$
ensemble is carried out using the Metropolis algorithm \citep{Metro53}
with bisection moves and displacement moves \citep{Chakravarty98}. The
total energy is calculated using the virial estimator
\citep{Herman82}.

The error estimate in the PIMC scheme is commonly given in powers of
the imaginary time time-step $\tau$ \citep{Ceperley95}. Therefore, in
order to systematically determine the thermal effects on the system we
have carried out all the simulations with $\tau = 0.03
E_\text{H}^{-1}$, where $E_\text{H}$ denotes the unit of Hartree.
Thus, the temperatures and the Trotter number $M$ are related by
$T=(k_{\text{B}}M\tau)^{-1}$, where $k_{\text{B}}$ is the Boltzmann
constant.

In the following we mainly use the atomic units, where the lengths,
energies and masses are given in the units of Bohr radius ($a_0$),
Hartree ($E_\text{H}$) and free electron mass ($m_\text{e}$),
respectively.  Thus, for the mass of the electrons we take
$m_\text{e}=1$ and for the protons $m_\text{p} = 1.83615267248\times
10^{3}m_\text{e}$. Conversion of the units of energy is given by
$E_\text{H} = 219474.6313705\text{cm}^{-1} \approx 27.2$ eV, and
correspondingly, $k_{\text{B}} = 3.1668152\times 10^{-6} E_\text{H}
\rm{K}^{-1}$.

The statistical standard error of the mean (SEM) with $2\, $SEM limits is
used as an error estimate for the evaluated observables.

For the $NVT$ simulations we place one H$_3^+$ ion, i.e.~three
protons and two electrons, into a cubic box and apply periodic
boundary conditions and the minimum image principle. The simulations
are performed in three different super cell (box) volumes:
$(300a_0)^3$, $(100a_0)^3$ and $(50a_0)^3$. These correspond to the
mass densities of $\sim 1.255\times 10^{-6}~\rm{gcm}^{-3}$, $\sim
3.388\times 10^{-5}~\rm{gcm}^{-3}$ and $\sim 2.710\times
10^{-4}~\rm{gcm}^{-3}$, respectively.  The density has no essential
effect at low $T$, where dissociation rarely takes place.  At higher
$T$, however, the finite density gives rise to the molecular
recombination balancing the more frequent dissociation.

It should be pointed out that application of the minimum image
principle with only one molecular ion in the periodic super cell may
both give rise to the finite-size effects and also disregard higher
density distribution effects, i.e.~fragments of several ions in the
simulation box.  Thus, the lower the density the better we minimize
the finite-size effects, which in this work are negligible, if not
absent. The zero density limit cannot be reached due to the finite
$T$.  To avoid all these ambiguities we define our targets as molecular
energetics, molecular partition function and other related molecular
quantities.  Therefore, in the following, we also exclude the trivial
contribution from the center-of-mass thermal dynamics and energy
$\tfrac{3}{2}\kB T$ to the molecular quantities.

The nuclear quantum dynamics, which was shown to be essential at low
$T$, turns out to be negligible at higher temperatures. It is
included, however, to be consistent with the low temperature results
and our earlier study.  For more details about the model and a
discussion about the here neglected contribution from the exchange
interaction see \cite{Kylanpaa10jcp}.

\section{Results and discussion}

\begin{figure}[t]
  \epsscale{1.1}
  \plotone{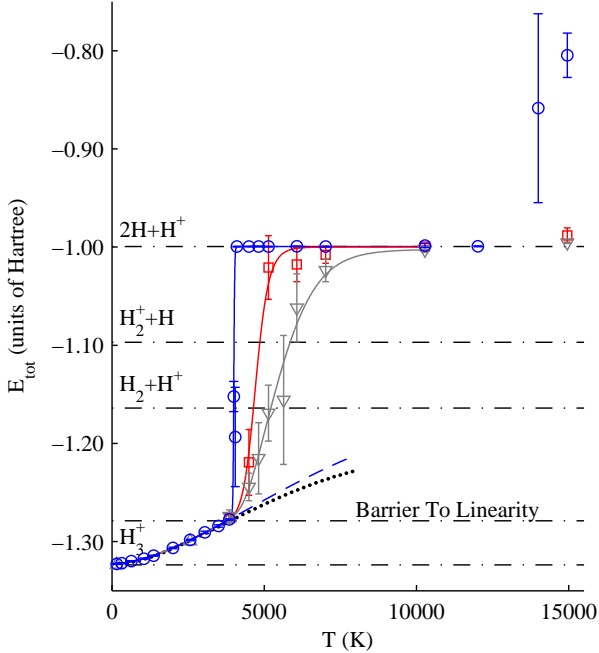}
  \caption{\label{Fig1}(Color online) $NVT$ total energy of the
     H$_3^+$ molecular ion as a function of temperature at three
     different densities: blue circles ($\sim 1.255\times
     10^{-6}~\rm{gcm}^{-3}$), red squares ($\sim 3.388\times
     10^{-5}~\rm{gcm}^{-3}$) and gray triangles ($\sim 2.710\times
     10^{-4}~\rm{gcm}^{-3}$). The blue dashed line is the energy
     fitted to Eq.~\eqref{eq:E3}. The black dots give the energy
     computed using the partition function fit given by
     \cite{Neale95ApJ}. The horizontal dash-dotted lines are the
     nonadiabatic zero Kelvin energies for the ion, its fragments and
     the barrier to linearity. The high temperature solid lines are
     mainly for guiding the eye, but used for numerical evaluation of
     the partition function, later.}
\end{figure}

\subsection{Overview of molecular energetics}

In Fig.~\ref{Fig1} the $NVT$ total energy (canonical ensemble internal
energy) of the H$_3^+$ ion and its fragments is shown as a function of
temperature. The molecular energy does not include the center-of-mass
translational kinetic energy $\tfrac{3}{2}\kB T$.  The data from
simulations are given as circles, squares and triangles corresponding
to the three densities. The PIMC data is also given in Tables
\ref{Table1} and \ref{Table2}.

The solid lines at $T < 4000$ K are fitted to analytical model forms
but at higher temperatures lines are for guiding the eye, only.  Our
low temperature fit and analytical model, Eq.~\eqref{eq:E3}, is given
as a blue dashed line and is discussed in the next sections in more
detail. For comparison the energies from the fitted partition
function of \cite{Neale95ApJ} is shown as black dots.  These two do
not manifest dissociation, and therefore, are not relevant at "higher
$T$".

The horizontal dash-dotted lines show the zero Kelvin energies for the
ion and its fragments in Eq.\eqref{fragments}. One of these lines
presents the energy for the "barrier to linearity", i.e.~the energy
needed for the transformation to linear molecular geometry.  It is
also seen to be roughly the barrier to dissociation within the
considered molecular densities.

Above $4000$ K the density dependence is clearly seen as varying
composition of fragments.  In the range from $4000$ to $10000$ K the
changing dissociation recombination balance leads to distinctly
different energetics, and above that, at our highest simulation
temperatures the thermal ionization of hydrogen atoms starts
contributing to the energy.  However, it is worth pointing out that
the temperature limits of these three ranges, i.e.~$0-4000$ K,
$4000-10000$ K and above $10000$ K, are subject to changes with larger
variation of densities.

Above $10000$ K in our lowest density case the thermal ionization of H
atoms is evident, see Fig.~\ref{Fig1}, but for our higher density
cases some $15000$ K is needed to show first signs of ionization.
Similar trend for the ionization is stated in \cite{Koskinen10ApJ},
although there the density is notably less than our lowest one.

\begin{figure}[t]
  \epsscale{1.1}
   \plotone{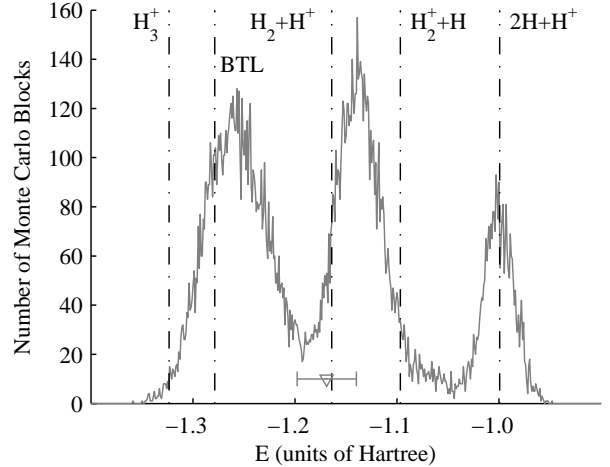}
   \caption{\label{Fig2}(Color online) Histogram of total energy
     sampling pinned in boxes of width $0.001E_\text{H}$ 
     from $(2\times 10^4) \times 10^5$ Monte Carlo samples
     averaged over blocks of $10^5$ samples.
     The simulation density, temperature and Trotter
     number are $\sim 2.710\times 10^{-4}~\rm{gcm}^{-3}$, $\sim
     5139.6$ K and $2048$, respectively. 
     Other notations are taken from Fig.~\ref{Fig1}.}
\end{figure}

Let us now consider the dissociation recombination reaction chain,
Eq.~\eqref{fragments}, and the contributing fragments to the quantum
statistical $NVT$ equilibrium trying to give an intuitive
classical-like picture of the composition.  With finite $T$, instead
of zero, we have finite $\beta$, instead of infinite, that brings
classical nature to the system the more, the higher the temperature.
In other words, the partial decoherence in our five particle
quantum system increases with increasing temperature, that
enables us to distinguish the fragments as separate molecules and
atoms in thermal equilibrium.
Based on this interpretation, we show the total energy distribution in
Fig.~\ref{Fig2} from sampling the imaginary time paths at about $5000$
K with $M=2048$ in our highest simulation density.

We see three main peaks  and by inspection
of the energy distribution the first and the second
can clearly be assigned to the rovibrationally excited
H$_3^+$ and H$_2+$H$^+$, respectively. As expected, there are no
rovibrational excitations available for $2$H$+$H$^+$,
leading to the peak average position very close to $-1E_\text{H}$. The
fourth fragment, H$_2^++$H, can be identified as the small
high-energy side shoulder of H$_2+$H$^+$ peak.  With the
interpretation of the area under the peak as the abundance of the
fragment in the equilibrium we find this contribution to be much
smaller than that of the others.

It is important to note, however, that the above illustration is
dependent on the block averaging procedure, see the caption of
Fig.~\ref{Fig2}. Pinning the energy data of each and every sample,
i.e.~choosing block of size one sample, would broaden the peaks in
Fig.~\ref{Fig2}. At the opposite limit, all samples in one block, the
single mean energy or ensemble average is the quantum statistical
expectation value $-1.169(29)E_\text{H}$, see Table \ref{Table2},
Figs.~\ref{Fig1} and \ref{Fig2}, where the statistical uncertainty
decreases with increasing simulation length.

\subsection{Molecular partition function}

To compare with the other published approaches for the molecular
partition function based on single molecule quantum chemistry we start
from the lowest temperature range from $0$ to $\sim 4000$ K, where the
molecule does not essentially dissociate, yet.

We present a low temperature H$_3^+$ molecular partition function as a
first approximation for the modeling of low density H$_3^+$ ion
containing atmospheres. Our aim is to find a simple analytical form,
which can be accurately fitted to the $NVT$ energies from our
simulations.

The partition function in terms of the Helmholtz free energy 
$F$ is written as
\begin{align}
\label{eqZ}
Z=e^{-\beta F},
\end{align}
where $\beta = (k_\text{B}T)^{-1}$, and the energy expectation value
is straightforwardly derived from the partition function as
\begin{align}
\label{eqE}
\ka{E} = -\frac{1}{Z}\frac{\partial Z}{\partial\beta}.
\end{align}
After solving the free energy from Eq.~\eqref{eqZ} as
\begin{align}
\label{FreeE}
F(T) = -k_\text{B}T \ln Z(T)
\end{align}
we write $F(T) = -k_\text{B}T f(T)$ and the
energy expectation value may be written as
\begin{align}
  \label{eq:E}
  \ka{E} = k_\text{B} T^2 \frac{\partial f(T)}{\partial T}.
\end{align}
We find that a well-behaving function fitting perfectly into our
simulation data,
\begin{align}
  \label{eq:E3}
  \ka{E} = k_\text{B} T^2 \left( ae^{-bT} + c \right) + de^{-\alpha/T},
\end{align}
allows analytical integration of Eq.~\eqref{eq:E} for $f(T)$ or $\ln
Z(T)$,
\begin{align}
  \label{eq:lnZ}
  \ln Z(T) = -\frac{a}{b} e^{-bT} + cT 
  + \frac{d}{k_\text{B}\alpha}e^{-\alpha/T} 
  + D.
\end{align}
Using the boundary condition for the molecular partition function with
a nondegenerate ground state, $Z(0) = 1$ or $\ln Z(0) = 0$, we get
$D=a/b$ in our model. However, inclusion of the contributions from the
ground state spin degeneracy factor and the zero-point rotations would
give $Z(0)=\xi>1$, which would lead to $D=a/b+\ln\xi$, and thus, shift
the function $\ln Z$ by a constant, only.

The weighted least squares fit of the above energy function,
Eq.~\eqref{eq:E3}, to our data for temperatures up to about 
$3900$ K, see Table \ref{Table1}, gives the parameters
\begin{align}
a & =      0.00157426 	    \nonumber\\
b & =      0.000132273     \nonumber\\
c & =      -6.15622 \times 10^{-6} \nonumber\\
d & =      0.00157430 	   \nonumber\\
\alpha & = 269.410 	    \nonumber\\
D & = a/b \approx 11.9016. \nonumber
\end{align}
In the fit, in addition to the $(2\text{SEM})^{-2}$ weights, we force
the first derivative of the energy with respect to the temperature to
be monotonically increasing up to $3900$ K.  The fit extrapolates the
$0$ K energy to about $0.000549E_\text{H}$ above that of the
para-H$_3^+$, i.e.~it gives an excellent match within the statistical
error estimate.

\begin{figure}[t]
  \epsscale{1.1}
   \plotone{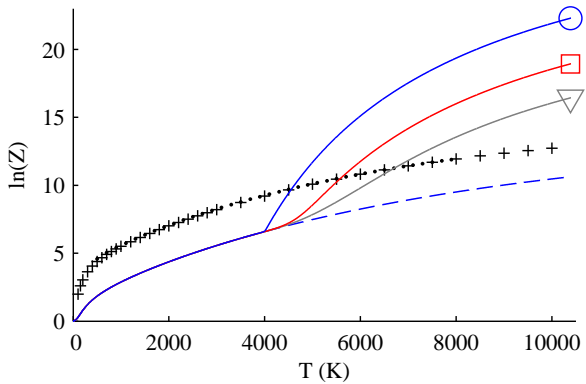}
   \caption{\label{Fig3}(Color online) The molecular 
    $NVT$ ensemble $\ln Z(T)$
   from the energetics in Fig.~\ref{Fig1} with the same notations.
   The blue solid line below $4000$ K and its extrapolation (dashed line)
   are from Eq.~\eqref{eq:lnZ}, whereas the curves for three
   densities are from Eq.~\eqref{eq:lnZhigh}.
   The black data (points) and fit (pluses) of
   \cite{Neale95ApJ} are also shown.}
\end{figure}

\begin{figure}[t]
  \epsscale{1.1}
   \plotone{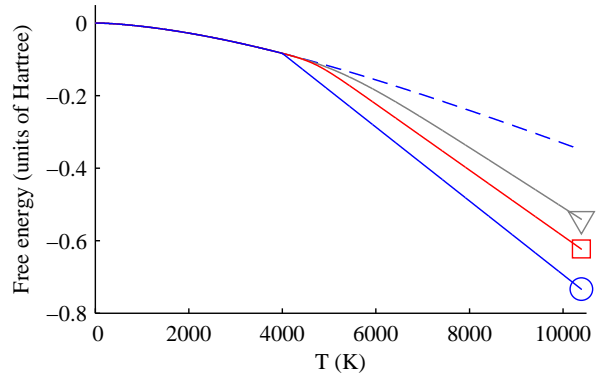}
   \caption{\label{Fig5}(Color online) Helmholtz free energy 
   from Eq.~\eqref{FreeE} (in units of Hartree). 
   Notations are the same as in Fig.~\ref{Fig3}.}
\end{figure}

\begin{figure}[t]
  \epsscale{1.1}
   \plotone{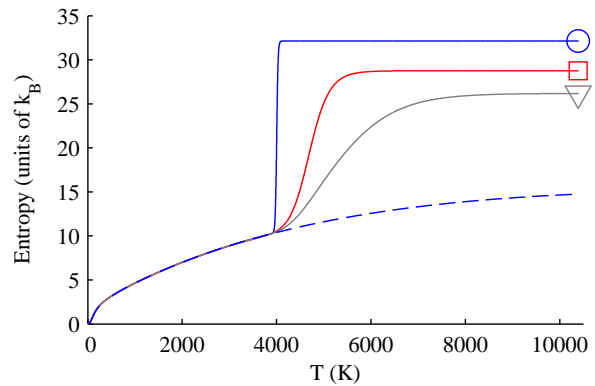}
   \caption{\label{Fig6}(Color online) Entropy 
   from Eq.~\eqref{entropy} (in units of $k_\text{B}$). 
   Notations are the same as in Fig.~\ref{Fig3}.}
\end{figure}

In Fig.~\ref{Fig3} the function $\ln Z(T)$ from Eq.~\eqref{eq:lnZ} is
shown in the range $0 < T < 4000$ K --- the behaviour of the model at
higher $T$ is illustrated by the dashed line.  Above $4000$ K the
three curves for different densities are obtained from those shown in
Fig.~\ref{Fig1} by numerical integration of Eq.~\eqref{eq:E} as
\begin{align}
 \label{eq:lnZhigh}
\ln Z(T) = \ln Z(T_1) + \int_{T_1}^{T} 
\frac{\ka{E}}{k_\text{B} T^2} ~\text{d} T,
\end{align}
where $T_1 = 500$ K.

\cite{Neale95ApJ} have presented the partition function $\ln Z(T)$
based on a semi-empirical potential energy surface, see
Fig.~\ref{Fig3}.  The overall shape is similar to the one of ours.
However, the energy $\ka{E}$ evaluated from their fit tends to be
systematically lower than ours, although roughly within our 2 SEM
error limits.  Thus, the deviations are not visible in
Fig.~\ref{Fig1}. For the partition function the main difference is in
their zero reference, which also leads to $\ln\xi = -\infty$.  This
difference in zero reference goes back to that of energetics: our
$\ka{E}$ in Eq.~\eqref{eq:E3} at $T = 0$ fits perfectly the zero
Kelvin energy of para-H$_3^+$, as mentioned earlier. The zero Kelvin
energy of the \cite{Neale95ApJ} behind their partition function is
that of the spin forbidden $J=0$ state, i.e.~$Z(0)=0$.  This in fact,
leads to the divergence of $\ln Z(T)$ at $T = 0$.

Our low temperature partition function Eq.~\eqref{eq:lnZ} is close to
complete.  With the PIMC approach we implicitly include all of the
quantum states in the system with correct weight without any
approximations.  This partition function is the best one for the
modeling of the low density H$_3^+$ ion containing atmospheres, at the
moment.  However, it is valid up to about $4000$ K, only.  As soon as
the density dependence starts playing larger role more complex models
are needed.  Such models can be fitted to our PIMC data given in
Tables \ref{Table1} and \ref{Table2}.

 \subsection{Other thermodynamic functions}

In Fig.~\ref{Fig5} we show the Helmholtz free energy from
Eqs.~\eqref{FreeE} and \eqref{eq:lnZ}, combined.  As expected, lower
density or larger volume per molecule lowers the free energy due to
the increasing entropic factor.  Dissociation and the consequent
fragments help in filling both the space and phase space more
uniformly or in less localized manner.

This kind of decreasing order is seen more clearly in the increasing
entropy, shown in Fig.~\ref{Fig6}.  The entropy has been evaluated
from
\begin{align}
 \label{entropy}
S = \frac{U-F}{T},
\end{align}
where the internal energy is $U = \ka{E} -\ka{E}_{T=0}$.  As expected,
both the total energy (internal energy) and entropy reveal the
dissociation similarly.

\begin{figure}[t]
  \plotone{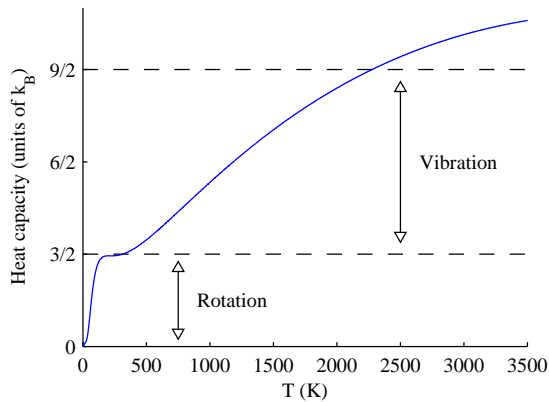}
   \caption{\label{Fig4}(Color online) Heat capacity as a function of
   temperature calculated using the analytical model of this work.
   The values on the y-axis are given in units of the Boltzmann
   constant $k_\text{B}$.}
\end{figure}

Finally, in Fig.~\ref{Fig4} we present the heat capacity
\begin{align}
\label{Cv}
C_V = \frac{\partial \ka{E}}{\partial T},
\end{align}
where $\ka{E}$ is taken from Eq.~\eqref{eq:E3}, which is valid at low
temperatures, only.

Considering the goodness of our functional form for $\ka{E}$, it is
very convincing to see the plateau at about $3/2 ~k_\text{B}$
corresponding to "saturation" of the contribution from the three
rotational degrees of freedom. Thus, above $200$ K the rotational
degrees of freedom obey the classical equipartition principle of
energy. It is the last term in the functional form of
Eq.~\eqref{eq:E3}, that gives the flexibility for such detailed
description of the energetics.

It should be emphasized that the plateau is not artificially
constructed to appear at $3/2~k_\text{B}$, except for a restriction
given for the first derivative of the total energy to be increasing.
Thus, the analytical model we present, Eq.~\eqref{eq:E3}, is found to
be exceptionally successful at low temperatures.

\renewcommand{\thefootnote}{\alph{footnote}}
\begin{table}[t]
  \begin{center}
    \caption{\label{Table1} $NVT$ energetics of the H$_3^+$ molecular
      ion at low temperatures --- here the same data applies for all
      three densities. The energies are given in the units of Hartree
      (atomic units) and with $2 \,$SEM error estimates.  The energies
      from our low $T$ fit (LTFIT) from Eq.~\eqref{eq:E3} and those
      from the fit of \cite{Neale95ApJ} (NT) are also given as
      comparison. At $0$ K the best upper bound is given, see the
      footnote\tablenotemark{c}.}
\begin{tabular}{cccc}
  \hline\hline
  $T (K)$ & PIMC\tablenotemark{a} & LTFIT\tablenotemark{a} & NT fit\tablenotemark{b}\\
  \hline
          $0$     &        &    $-1.3231$   &     $(-1.32367)$\tablenotemark{c}
    \\
       $\sim 160.61$   &   $-1.3227(7) $  &    $-1.3227$   &    $-1.3232$\\
       $\sim 321.22$   &   $-1.3221(6) $  &    $-1.3220$   &    $-1.3225$\\
       $\sim 642.45$   &   $-1.3198(6) $  &    $-1.3202$   &    $-1.3209$\\
       $\sim 1052.6$   &   $-1.3173(7) $  &    $-1.3171$   &    $-1.3179$\\
       $\sim 1365.2$   &   $-1.3143(5) $  &    $-1.3141$   &    $-1.3148$\\
       $\sim 2000.3$   &   $-1.3064(7) $  &    $-1.3065$   &    $-1.3070$\\
       $\sim 2569.8$   &   $-1.2983(8) $  &    $-1.2984$   &    $-1.2989$\\
       $\sim 3049.2$   &   $-1.2905(12)$  &    $-1.2909$   &    $-1.2917$\\
       $\sim 3499.3$   &   $-1.2840(12)$  &    $-1.2835$   &    $-1.2847$\\
       $\sim 3855.6$   &   $-1.2774(7) $  &    $-1.2774$   &    $-1.2792$\\
\hline
\end{tabular}
\footnotetext[1]{This work.}
\footnotetext[2]{Calculated from the fit given in \cite{Neale95ApJ}.}
\footnotetext[3]{Para-H$_3^+$, see refs.~\cite{Jaquet06} and\\ \cite{Adamowicz09jcp1}.} 
\end{center}
\end{table}
\renewcommand{\thefootnote}{\arabic{footnote}}

\begin{table}[t]
\begin{center}
  \caption{\label{Table2} PIMC $NVT$ energetics of the H$_3^+$ molecular
    ion at high temperatures for the three densities
    (expressed as the number of molecular ions per volume),
    see Fig.~\ref{Fig1}. Notations are the same as in Table \ref{Table2}.}
  \begin{tabular}{cccc}
    \hline\hline
    $T (K)$ & $(300a_0)^{-3}$ & $(100a_0)^{-3}$ & $(50a_0)^{-3}$\\
    \hline
       $\sim 3999.2$  &  $-1.152(16)$ &               &               \\ 
       $\sim 4050.0$  &  $-1.19(6) $  &		 &		 \\
       $\sim 4100.4$  &  $-0.9995(4)$ &		 &		 \\
       $\sim 4498.2$  &  $-0.9993(4)$ &  $-1.219(34)$ &  $-1.244(15)$ \\
       $\sim 4819.5$  &  $-0.9993(4)$ &               &  $-1.215(37)$ \\
       $\sim 5139.6$  &  $-0.9995(4)$ &  $-1.020(33)$ &  $-1.169(29)$ \\
       $\sim 5634.8$  &               &               &  $-1.156(66)$ \\
       $\sim 6070.3$  &  $-0.9991(4)$ &  $-1.018(18)$ &  $-1.062(35)$ \\
       $\sim 7017.2$  &  $-0.9995(4)$ &  $-1.008(9)$  &  $-1.024(12)$ \\
       $\sim 10279$   &  $-0.997(3)$  &  $-0.9995(8)$ &  $-1.003(3)$  \\
       $\sim 12016$   &  $-0.9993(6)$ &               &               \\
       $\sim 13997$   &  $-0.86(10)$  & 	      &               \\
       $\sim 14951$   &  $-0.805(23)$ &  $-0.988(8)$  &  $-0.9957(8)$ \\
\hline
\end{tabular}
\end{center}
\end{table}

\section{Conclusions}
We have evaluated the temperature dependent quantum statistics of the
five particle molecular ion H$_3^+$ far beyond its dissociation
temperature at about $4000$ K.  This is done with the path integral
Monte Carlo (PIMC) method, which is basis set and trial wave function
free approach and includes the Coulomb interactions exactly.  Thus, we
are able to extend the traditional {\it ab initio} quantum chemistry
with full account of correlations to finite temperatures without any
approximations, also including the nuclear thermal and quantum
dynamics.

It is fair to admit, however, that PIMC is computationally heavy for
good statistical accuracy and approximations are needed to solve the
"Fermion sign problem" in cases where exchange interaction becomes
essential.

The temperature dependent mixed state description of the H$_3^+$ ion,
the density dependent equilibrium dissociation recombination balance
and the energetics has been evaluated for the first time.  With the
rising temperature the rovibrational excitations contribute to the
energetics, as expected, whereas the electronic part remains in its
ground state in the spirit of the Born--Oppenheimer approximation. At
about $4000$ K the fragments of the molecule, H$_2+$H$^+$, H$_2^++$H
and $2$H$+$H$^+$, start contributing. Therefore, H$_3^+$ ion becomes
less dominant, and eventually negligible in high enough $T$.

We have shown how the partial decoherence in the mixed state can be
used for interpretation of the fragment composition of the equilibrium
reaction.  Furthermore, we have evaluated explicitly the related
partition function, free energy, entropy and heat capacity, all as
functions of temperature.  An accurate analytical functional forms are
given for the temperatures below dissociation.  We consider all these
as major improvements to the earlier published studies, where
dissociation has not been considered.

\section{Acknowlegements}

For financial support we thank the Academy of Finland, and for
computational resources the facilities of Finnish IT Center for
Science (CSC) and Material Sciences National Grid Infrastructure
(M-grid, akaatti).

\end{document}